%% file: moriond.tex
\documentclass[11pt]{article}
\usepackage{moriond,epsfig,cite,multirow}
\usepackage{amsmath}

\input{babarsym}
\input{mysym}

\begin{document}
\input{title_page}
\input{introduction}
%
\input{radiative}
\input{CP_bsg}
\input{Kstar_gamma}
\input{Kstar_gamma_CP}
\input{Kstar2_gamma}
%
\input{bsll}
\input{Bmunu}

\section{Conclusions}
The unprecedented luminosity of the \B-factories permitted to perform
new extensive and detailed tests on the processes
$\b \to \s \gamma$ and $\b \to \s \ell^+ \ell^-$ 
involving flavour changing neutral currents. There are no experimental
evidences of \CP violation in the $\b \to \s \gamma$ at the $5\%$ level
and the SM predictions are confirmed. Both \babar and Belle are
collecting a richer data sample that will consent more stringent tests
of these aspect of the SM and eventually bring in a near future some 
surprises. 
\section*{References}

\bibliography{moriond}
\end{document}

%% file: mysym.tex
\def\Br  {\ensuremath{\mathcal {B.R.} }\xspace}
\def\acp {\ensuremath{\mathcal{A}_{\CP}}\xspace}

%% file: title_page.tex
\vspace*{4cm}
\title{Radiative Penguin and Leptonic Rare Decays at \babar}

\author{ Eugenio Paoloni }

\address{Dipartimento di Fisica dell'Universit\`a degli studi di Pisa\\
 e\\
Istituto Nazionale di Fisica Nucleare, Sez. di Pisa\\
via Buonarroti 2, 56100 Pisa, Italy\\
(for the {\babar}\ collaboration)
}

\maketitle\abstracts{
  Recent {\babar}\  results on rare \B decays involving flavour-changing 
  neutral
  currents or  purely leptonic final states are presented.
  New measurements of the \CP asymmetries in
  $\B \rightarrow \Kstar \gamma$, 
  $\B \rightarrow \Kstar_2(1430) \gamma$, and 
  $\b \rightarrow \s \gamma$ are reported,
  as well as a new measurement of the 
  $\B \rightarrow \Kstar \gamma$ branching fraction.
  Also reported are updated limits on 
  $\B \rightarrow \mu \nu$ and recent measurements of
  $\B \rightarrow  K^{(*)}\ell \ell$  and $\b \rightarrow \s \ell \ell$.
  The data sample comprises $123 \cdot 10^6$ $\Y4S \rightarrow \B\Bbar$
  decays collected with the \babar\ detector at the \pep2 \epem storage ring.
}
\vspace{4cm}
\begin{center}
\hrule
Contributed to the Proceedings of the {\sc XXXIX} Recontres de Moriond
Electroweak Interactions and Unified Theories
\end{center}
\newpage

%% file: introduction.tex
\section {Introduction}

The study of radiative and leptonic rare \B decays in search for 
significant discrepancies with respect to the Standard Model (SM) 
theoretical predictions represent a very attractive field of research. 
These decays can show some of the visible effects predicted by
many extensions of the SM whose measure allows to constraint 
(if not potentially to discover) new physics.

The \babar\ collaboration exploited the unprecedented \pep2 luminosity
to perform an extensive and detailed series of studies on these decays
analysing a sample that comprise $123 \cdot 10^6$ $\Y4S \rightarrow \B\Bbar$
decays recorded by the \babar\ detector\cite{Aubert:2001tu}.

This paper will summarize the results of these searches.


%% file: radiative.tex
\section{\CP violation in radiative \B Decays}
Radiative decays $\b\rightarrow \s \gamma$ proceed at leading order
in the SM through one loop penguin diagrams. 
The new fields predicted by many extensions of the SM can contribute
with additional amplitudes to this process appearing as virtual particles 
in the penguin loop diagrams. 
The comparison of the measured inclusive branching ratio (world average 
$\Br (\B \rightarrow X_\s \gamma)= (3.3 \pm 0.4) \cdot 10^{-4}$
 \cite{Hagiwara:2002fs}) with respect to the SM theoretical
predictions ( $(3.6\pm 0.3)\cdot 10^{-4}$ \cite{Gambino:2001ew,Buras:2002tp} )
has already provided some constraints on the new physics beyond the SM 
\cite{Hurth:2003vb}. 

The measurement of the \CP violation can shed new
light on the structure of this flavour changing neutral current both
testing the SM predictions and constraining the space parameter of 
the SM extensions.


%% file: CP_bsg.tex
\begin{figure}
\centering{
\includegraphics[width=15cm]{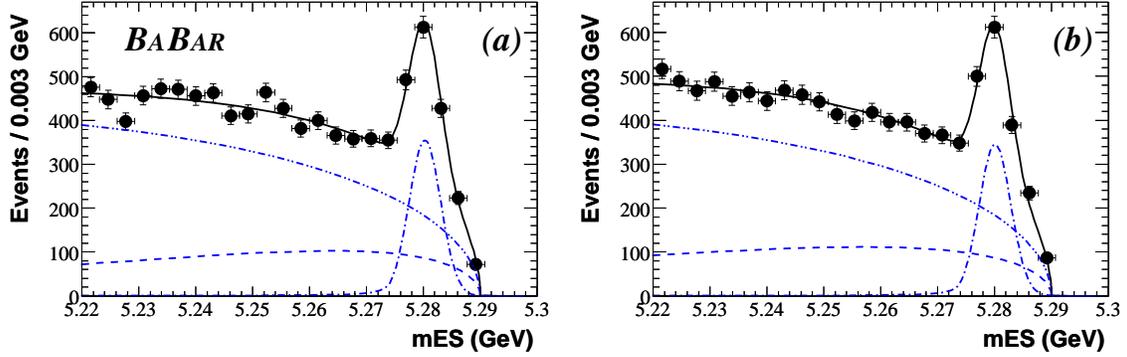}}
\caption{Fits to the beam--energy substituted mass distributions in 
data events for:
(a) $\Bbar\to X_\s\gamma$, (b) $\B\to X_{\sbar}\gamma$
Contributions are shown from peaking Crystal Ball (dotted--dashed), 
fixed continuum Argus shape (dotted) and free \BB\ and cross--feed 
Argus shape (dashed).}
\label{fig_CP_yields}
\end{figure}
\subsection{ Direct \CP violation in $\Bbar \rightarrow X_\s \gamma$ decays}
In the SM the \CP violation in the inclusive process 
$\Bbar \rightarrow X_\s \gamma$ can be reliably predicted \cite{Hurth:2003pn}:
\begin{equation}
\acp = \frac{ \Gamma\left( \Bbar \rightarrow X_\s\, \gamma \right) \, - \,
        \Gamma\left( \B \rightarrow X_{\sbar}\, \gamma \right)}
{ \Gamma\left( \Bbar \rightarrow X_{\s}\, \gamma\right) \, + \,
        \Gamma\left( \B \rightarrow X_{\sbar}\, \gamma\right)}
        \, = \, 0.0044^{+0.0024}_{-0.0014}
\end{equation}
whereas in some Super Symmetric scenarios sizable asymmetries
 ( $\acp \sim 10\%$ ) are possible and natural\cite{Kagan:1998bh}.

In this analysis \cite{Aubert:2004hq}
a sample of $(88.9\pm 1.0)\times 10^6$ \BB\ 
pairs collected at the \Y4S\ resonance is used.
The $\Bbar \to X_\s \gamma$ sample is obtained combining the
twelve full reconstructed self-tagging decay channels: 
\begin{alignat*}{3}
\Bub &\to& \Km\piz\gamma, \Km\pip\pim\gamma, \Km\piz\piz\gamma, 
\Km\pip\pim\piz\gamma\\
\Bzb &\to& \Km\pip\gamma, \Km\pip\piz\gamma, \Km\pip\piz\piz\gamma, \Km\pip\pim\pip\gamma\\
\Bub &\to& \KS\pim\gamma, \KS\pim\piz\gamma, \KS\pim\piz\piz\gamma, \KS\pim\pip\pim\gamma.
\end{alignat*}
Their charge conjugate is used to obtain the $\B \to X_{\sbar} \gamma$ 
sample. 
Fully reconstructed  $\B \to X_{\sbar}\gamma$ decays are characterized 
by two kinematic variables: the beam--energy substituted mass, 
$m_{ES}= \sqrt{(\sqrt{s}/2)^2- p^{*2}_\B}$, and the energy difference 
between the \B\ candidate and the beam--energy,
$\Delta E = E_\B^* - (\sqrt{s}/2)$, where $E_\B^*$ and $p_\B^*$ are the 
energy and momentum of the \B\ candidate in the $e^+e^-$ center--of--mass 
frame, and $\sqrt{s}$ is the total center--of--mass energy.
We require candidates to have $|\Delta E|<0.10$~\gev, and then we fit the $m_{ES}$
distribution between 5.22 and 5.29~\gev\ to extract the signal yield.
The positive identification of charged kaons removes any contribution from 
$b\to d\gamma$.

\acp is obtained from the yield asymmetry between the \B and the \Bb
sample correcting for flavour misidentification and detector asymmetry.

In Figure~\ref{fig_CP_yields} the final fits to the 
$m_{ES}$ distributions for $\b\to s\gamma$ and $\bbar\to\sbar\gamma$ 
events are presented. 

We measure an \CP asymmetry of $(0.025\pm 0.050\pm 0.015)$, where the 
first errors is statistical and the second systematic, corresponding 
to an allowed range of $-0.06  < \acp (\b\to s\gamma)< +0.11$ at 90\%\ 
confidence level in good agreement with the SM predictions.


%% file: Kstar_gamma.tex
\begin{figure}[tb]
 \includegraphics[width=\linewidth]{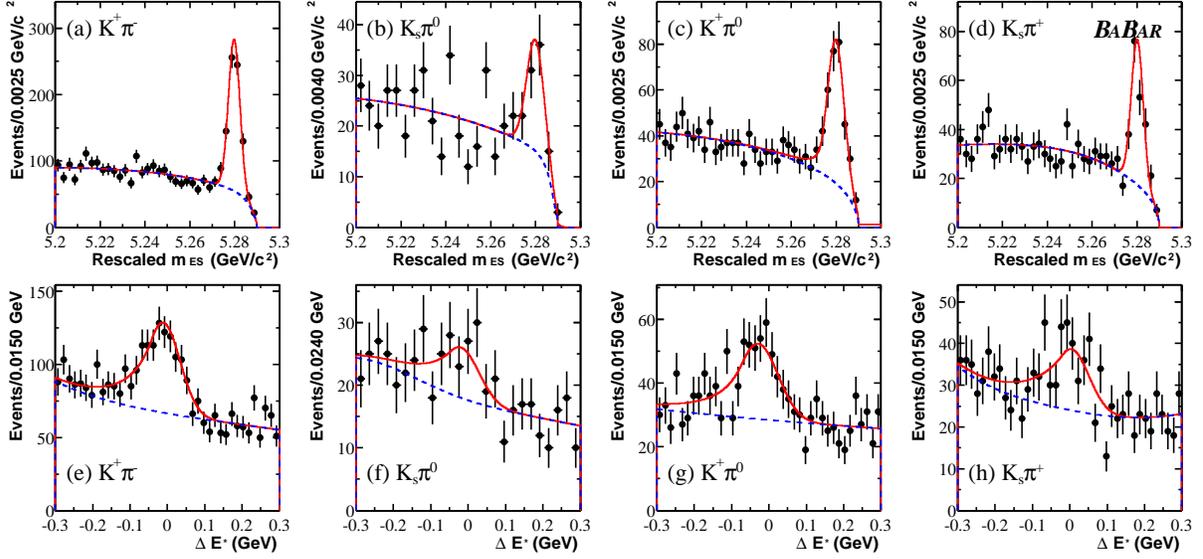}
\caption{$M_{ES}$ and $\Delta E^*$ distributions for the $\B \to \Kstar \gamma$
 candidates. The solid and dashed curves show respectively the projections 
of the complete fit and background component alone.}
\label{fig:Kstar_mes}
\end{figure}
\begin{table*}[tb] 
  \label{table:kstar_gamma_results}
  \caption{$\B \to \Kstar \gamma$ analysis: the results from the likelihood fit are summarized. The signal
    efficiency $\epsilon$, the fitted signal yield $N_S$, the
    branching fractions \BR and the CP-asymmetries $\acp$ for each
    decay mode are shown.
    The combined branching fractions for
    $\Bz \to \Kstarz \gamma$ or $\Bp \to K^{*+}\gamma$ are also shown.
    Errors are statistical and systematic, respectively, with
    the exception of $\epsilon$ and $N_S$ where $\epsilon$ only has a
    systematic error $N_S$ has only a
    statistical error.}
      
      
      
      
      
        \begin{center}
        \begin{tabular*}{\linewidth}{@{\extracolsep{\fill}}lr@{\extracolsep{0pt}}@{.}l@{\extracolsep{\fill}}r@{\extracolsep{0pt}}@{.}l@{\extracolsep{\fill}}r@{\extracolsep{0pt}}@{.}l@{\extracolsep{\fill}}r@{\extracolsep{0pt}}l@{\extracolsep{\fill}}r@{\extracolsep{0pt}}@{.}l}
\hline\hline

    Mode&
    \multicolumn{2}{c}{$\epsilon$(\%)}&
    \multicolumn{2}{c}{$N_S$}&
    \multicolumn{2}{c}{${\cal B}\times\mbox{10}^{-\mbox{5}} $}&
    \multicolumn{2}{c}{combined ${\cal B}\times\mbox{10}^{-\mbox{5}} $}&
    \multicolumn{2}{c}{${\acp}$}\\\hline

    $K^+\pi^-$&
    24&
    4$\pm$1.4&
    582&        6$\pm$29.7&
    \ \ 3.92$\pm$0& 20$\pm$0.23&
    \multirow{2}*{{\Big \}}\ 3.92$\pm$0.}& \multirow{2}*{20$\pm$0.24}&
    -0&    069$\pm$0.046$\pm$0.011\\

    $K_s\pi^0$&
    15&
    3$\pm$1.9&
    61&8$\pm$15.3&
    \ \ 4.02$\pm$0& 99$\pm$0.51&
        &&
    \multicolumn{2}{c}{}\\

    $K^+\pi^0$&
    17&
    4$\pm$1.6&
    250&9$\pm$22.6&
    \ \ 4.90$\pm$0& 45$\pm$0.46&
    \multirow{2}*{{\Big\}}\ 3.87$\pm$0.}& \multirow{2}*{28$\pm$0.26}&
    0& 084$\pm$0.075$\pm$0.007\\

    $K_s\pi^+$&
    22&
    1$\pm$1.4&
    156&
    9$\pm$15.7&
    \ \ 3.52$\pm$0& 35$\pm$0.22&
        &&
    0&    061$\pm$0.092$\pm$0.006\\\hline\hline

    \end{tabular*}
    \end{center}

\end{table*}

\subsection{ Search for  \CP or isospin asymmetries in the $\B \rightarrow \Kstar \gamma$ decays}
The set of exclusive decays $\B \to \Kstar \gamma$ provide other opportunities to 
test the SM predictions for the isospin ($\Delta_{0-}$, Eq.~\ref{eq:Kstar_isospin})  
and the \CP asymmetries ($\acp$, Eq.~\ref{eq:Kstar_acp}):
\begin{eqnarray}
\Delta_{0-} = \frac{\Gamma(\overline{B^0} \to \overline{K^{*0}}\gamma) 
- \Gamma(B^{-}\to K^{*-}\gamma)}
{\Gamma(\overline{B^0} \to \overline{K^{*0}}\gamma) +
\Gamma(B^{-}\to K^{*-}\gamma)},
\label{eq:Kstar_isospin}\\
\acp = \frac{\Gamma(\overline{B} \to \overline{K^*}\gamma) - \Gamma(\B
\to \Kstar \gamma)}{\Gamma(\overline{B} \to \overline{K^*}\gamma)
+ \Gamma(\B \to \Kstar \gamma)}.
\label{eq:Kstar_acp}
\end{eqnarray}
The  SM  predicts a positive $\Delta_{0-}$ between $5$ and $10\%$
~\cite{Kagan:2001zk}, and $\acp$ less than $1\%$~\cite{Kagan:1998bh}.
New physics contributions can modify these values significantly.

This analysis uses a sample  of $\mbox{88.2}\times \mbox{10}^6 B\overline{B}$.
The \Kstar is reconstructed in the self-tagging decay channels
$\Kstarz \to \Kp \pim\; ; \; \Kstarp \to \Kp \piz\, , \, \KS\pip$ and 
their charge conjugates. For the isospin analysis was also used 
$\Kstarz \to \KS\piz$. 

The signal yield and \acp for each decay mode are reported in table
\ref{table:kstar_gamma_results}, they are determined with
a two-dimensional extended unbinned maximum likelihood fit
to the $m_{ES}$ and $\Delta E^*$.
$\Delta_{0-}$ is determined from the signal yields correcting for
differences in signal efficiency and lifetime among the neutral 
and charged \B. The preliminary results are:
\begin{alignat}{4}
      \acp        &= \, -0.015 & \pm\, 0.036 \, (\mathrm{stat.})\,&
      \pm0.010 \, (\mathrm{sys.})\,&\\
      \Delta_{0-} &= \, +0.051 & \pm\, 0.044 \, (\mathrm{stat.})\,&
      \pm\, 0.023 \, (\mathrm{sys.})\, &
      \pm\, 0.024 \, (R^{+/0})         
\end{alignat}
the first being the statistical and the second the systematic error.
The third error on $\Delta_{0-}$ is related to the uncertainty on
the ratio of the branching ratios recently measured by the \babar 
collaboration \cite{Aubert:2004ur}:
$$
R^{+/0} = 
\frac{\Br\left(\Y4S \rightarrow \Bz \Bzb\right)}
     {\Br\left(\Y4S \rightarrow \Bp \Bm\right)} = 1.006 \pm 0.048 
$$
and accounts for the possibility of different production rate of 
charged and neutral \B.
%

%% file: Kstar_gamma_CP.tex
\subsection{ Search for time dependent \CP asymmetry in 
$\B\rightarrow \Kstar \gamma \, (\Kstar \rightarrow \KS \piz)$}

The final state $\KS \piz \gamma$ is accessible to both the $\B$ and the 
$\Bb$ through $\Kz-\Kzb$ mixing:
\begin{alignat*}{3}
 \Bz  &\to& \Kstarzb \gamma &(\Kstarzb \to \KS \piz )\\
 \Bzb &\to& \Kstarz \gamma  &(\Kstarz \to \KS \piz ).
\end{alignat*}
The interference between decay and mixing can produce a time dependent
\CP asymmetry:
$$
    A_{\CP}(t) = 
    \frac {\Gamma (\Bz(t) \rightarrow \KS \piz \gamma) -
      \Gamma (\Bzb(t) \rightarrow \KS \piz \gamma) }
    {\Gamma (\Bz(t) \rightarrow \KS \piz \gamma) +
      \Gamma (\Bzb(t) \rightarrow \KS \piz \gamma) } = 
        S\, \sin \Delta m\, t \;-\; C    $$ 
where $t$ is the time elapsed since the \B meson production, and $C$ is the 
direct \CP asymmetry measured in the analysis previously described.
In the SM the helicity structure of the hadronic currents strongly suppress 
the time dependent \CP asymmetry \cite{Atwood:1997zr}:
$$ S \,=\, 2\, \frac{m_s}{m_b}\, \sin \, 2 \beta \sim 4\% \; , \; |C| < 1\% $$

We analyzed 124 million $\Y4S\to\BB$ decays selecting 1916 $\B \to \KS \gamma$
candidates. The decay point of these candidates is obtained intersecting
the $\KS$ flight direction with the beams line. The flavour and the decay 
point of the companion \B is determined with an inclusive reconstruction 
\cite{Aubert:2002rg}. 
The fit to $\acp(t)$ give the preliminary result\cite{Aubert:2004pe}:
\begin{alignat*}{3}
 S &=  0.25  &\pm 0.63 &\pm 0.14\\
 C &= -0.57  &\pm 0.32 &\pm 0.09
\end{alignat*}
where the first error is statistical and the second systematic.


%% file: Kstar2_gamma.tex
\subsection{Measurement of the $\B\rightarrow \Kstar_2(1430) \gamma$
branching ratio}
During the last year the \babar collaboration analyzed $88.5\times 10^6$ \BB 
events to measure the  $\B\rightarrow \Kstar_2(1430) \gamma$
branching ratio \cite{Aubert:2003zs}. The candidates are reconstructed in the 
decay channels:
\begin{alignat*}{3}
\Bzb &\to \Kstar_2(1430)^0\gamma\quad & ({\Kstar_2}^0 &\rightarrow \Kp\pim)\\
\Bp  &\to \Kstar_2(1430)^+\gamma\quad & ({\Kstar_2}^+ &\rightarrow\Kp\piz,
        \KS\pip)
\end{alignat*}
It is interesting to note that the most dangerous background in this analysis
is represented by other rare $\B \to X_\s  \gamma$ modes 
(mainly $\B\rightarrow \Kstar(1410)\gamma$ ). The $\Kstar_2(1430)$ is a 
$J^P = 2^+$ resonance whereas the $\Kstar(1410)$ has $J^P=1^-$ then the
distribution of the helicity angle $\vartheta_H$ of the kaon in the \Kstar
rest frame does differ for signal and background.
We measure the signal yield with a multidimensional fit to $\cos \vartheta_H$,
the energy substituted mass $m_{ES}$, and the center-of-mass frame energy 
difference $\Delta E$. The preliminary measurement of the branching fractions 
are:
\begin{alignat*}{6}
\Br \big( &\Bzb &\to \Kstar_2(1430)^0\gamma &\big)\, & =\, (1.22 &\pm0.25 &\pm0.11)\times10^{-5}\\
\Br \big( &\Bp  &\to \Kstar_2(1430)^+\gamma &\big)\, & =\, (1.44 &\pm0.40 &\pm0.13)\times10^{-5}
\end{alignat*}
in good agreement with the SM theoretical prediction \cite{Cheng:2004yj}:
$$
\Br \big( \B \to \Kstar_2(1430)\gamma \big)\,  =\, (1.48 \pm0.30)\times10^{-5}.
$$

%% file: bsll.tex
\section{Measurement of $\b \to \s \ell^+ \ell^-$ branching fractions }
The process $\b \to \s \ell^+ \ell^-$ is another very promising field of
research. With respect to the $\b \to \s \gamma$ decays there are two 
additional diagrams contributing to the process: a penguin diagram with
a \Z insertion and a box diagram:
$$
\includegraphics[height=4cm]{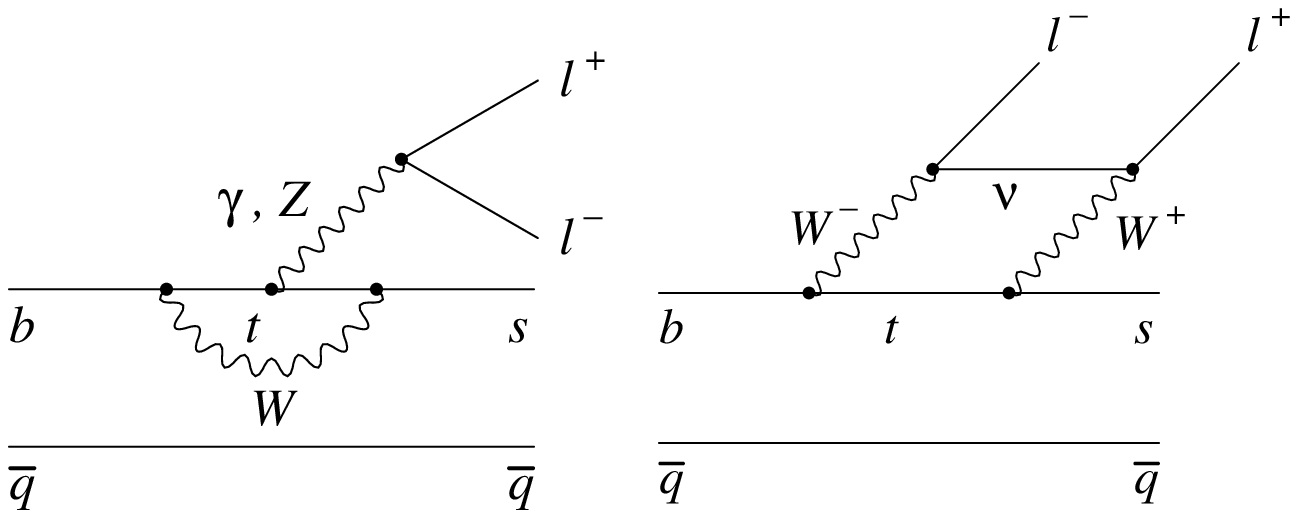}
$$
Since the last year the \babar collaboration finalized on $113 \invfb$
\cite{Aubert:2003cm} the measurement of the branching ratios 
$\B \to K^{(*)} \ell^+ \ell^-$.
\begin{alignat*}{5}
\Br \big(& \B \to K \ell^+ \ell^-  & \big) & = \big( 6.5^{+1.4}_{-1.3} & \pm 0.4 &\big) \times 
10^{-7}\\
\Br \big(& \B \to \Kstar \ell^+ \ell^- & \big) & = \big( 8.8^{+3.3}_{-2.9} & \pm1.0 & \big) \times 10^{-7},
\end{alignat*}
and performed a preliminary measurement of the inclusive branching 
ratio $\B \to X_\s \ell^+ \ell^-$ using a sum over 
exclusive modes in which $X_\s$ is composed by one kaon, one charged pion
and/or one neutral pion \cite{Aubert:2003rv}:
$$
\Br\left(\B \to X_\s \ell^+ \ell^- \right) = 
\left(6.3 \pm 1.6^{+1.8}_{-2.5} \right) \times 10^{-6}.
$$

All these results are in good agreement with the SM theoretical predictions.

%% file: Bmunu.tex
\section{Leptonic \B decays}
The study of the purely leptonic $\Bp \to \ell^+ \nu_\ell$ can provide
sensitivity to poorly constrained SM parameters and also act as a probe for
new physics. In the SM the reaction proceed through $\bbar \u \to \Wp \to\ell^+ \nu_\ell $ and the branching ratio is given by:
$$
\Br(\Bp \rightarrow \ell^+ \nu)\,=\,
    \frac{G_F^2\,m_{\B}\,m_\ell^2}{8 \pi}
    \left( 1- \frac{m_\ell^2}{m_{\B}^2}\right)^2\,f_{\B}
  \left|V_{ub}\right|^2 \tau_{\B},
$$
where $G_F$ is the Fermi coupling constant, $m_\ell$ and $m_\B$ are the
lepton and meson masses, $f_B$ is the \B decay constant, $V_{ub}$ is the 
relevant CKM matrix element and $\tau_\B$ is the \B+ lifetime.
Currently the $f_B$ parameter comes from lattice QCD simulation and is
affected by a $15\%$ uncertainty. Observation of $\Bp \to \ell^+ \nu_\ell$
could provide the first direct measurement of $f_B$.
Unfortunately leptonic decays are strongly suppressed by helicity and there
are not yet experimental evidence for such decays. Using $88.4$ million 
$\B\Bbar$ events \babar set an upper limit on the branching ratio
$$
\Br(\Bp \rightarrow \mup \nu_\mu)\,< 6.6 \cdot 10^{-6}
$$
at the $90\%$ confidence level. This limit is consistent with the SM predictions.
